\documentclass[journal]{IEEEtran}
\usepackage{graphicx}
\usepackage{mathrsfs}
\usepackage{amsmath}
\usepackage{amssymb}
\usepackage{amsthm}
\usepackage{lineno}
\usepackage{multirow}
\usepackage{booktabs}
\usepackage{algorithm}
\usepackage{algorithmic}
\usepackage{color}
\usepackage{multicol}
\usepackage{stfloats}
\usepackage{footmisc}
\usepackage{subfigure}
\usepackage[]{cite}

\usepackage{enumitem}
\usepackage{epstopdf}
\usepackage{setspace}
\makeatletter
\renewcommand{\maketag@@@}[1]{\hbox{\m@th\normalsize\normalfont#1}}%
\makeatother

\begin{document}

\title{Channel-Transferable Semantic Communications for Multi-User OFDM-NOMA Systems}

\author{Lan~Lin,
        Wenjun Xu*,~\IEEEmembership{Senior Member,~IEEE,}        
        Fengyu Wang,
        Yimeng Zhang,
        Wei Zhang,~\IEEEmembership{Fellow,~IEEE,}              
        \\
        Ping Zhang,~\IEEEmembership{Fellow,~IEEE}
        
\thanks{
This work was supported in part by the National Natural Science Foundation of China under Grant 62293485; in part by BUPT Excellent Ph.D. Students Foundation CX2022230. \textit{(Corresponding author: Wenjun Xu)}

Lan Lin and Yimeng Zhang are with the State Key Laboratory of Network and Switching Technology, Beijing University of Posts and Telecommunications, Beijing 100876, China (e-mail: aimo614@bupt.edu.cn, yimengzhang@bupt.edu.cn).
%

Wenjun Xu and Ping Zhang are with the State Key Laboratory of Network and Switching Technology, Beijing University of Posts and Telecommunications, Beijing 100876, China, and also with the Peng Cheng Laboratory, Shenzhen 518066, China (e-mail: wjxu@bupt.edu.cn, pzhang@bupt.edu.cn).

Fengyu Wang is with the School of Artificial Intelligence, Beijing University of Posts and Telecommunications, Beijing, 100876, China (e-mail:fengyu.wang@bupt.edu.cn). 

Wei Zhang is with the School of Electrical Engineering and Telecommunications, University of New South Wales, Sydney, NSW 2052, Australia (e-mail: w.zhang@unsw.edu.au).

}
}

\maketitle

\begin{abstract}
Semantic communications are expected to become the core new paradigms of the sixth generation (6G) wireless networks. 
Most existing works implicitly utilize channel information for codecs training, which leads to poor communications when channel type or statistical characteristics change. To tackle this issue posed by various channels, a novel channel-transferable semantic communications (CT-SemCom) framework is proposed, which adapts the codecs learned on one type of channel to other types of channels. Furthermore, integrating the proposed framework and the orthogonal frequency division multiplexing systems integrating non-orthogonal multiple access technologies, i.e., OFDM-NOMA systems, a power allocation problem to realize the transfer from additive white Gaussian noise (AWGN) channels to multi-subcarrier Rayleigh fading channels is formulated. We then design a semantics-similar dual transformation (SSDT) algorithm to derive analytical solutions with low complexity. Simulation results show that the proposed CT-SemCom framework with SSDT algorithm significantly outperforms the existing work w.r.t. channel transferability, e.g., the peak signal-to-noise ratio (PSNR) of image transmission improves by $4.2-7.3$~dB under different variances of Rayleigh fading channels.

\end{abstract}


\IEEEpeerreviewmaketitle

\section{Introduction}%
\IEEEPARstart {W}{ith} the emergence of a large number of intelligent applications in the sixth generation (6G) wireless communications\cite{intelligent_apps}, traditional communication systems are facing severe challenges and in desperate need of new breakthroughs. To address the challenges, semantic communications are considered to be efficient and promising technologies due to their concise properties \cite{engineering_ping}. Semantic communications, which focus on information connotation, aim to extract the semantic features of information for transmission and recover the information from the received semantic features. Thanks to the development of artificial intelligence (AI) and deep learning (DL), it has become possible to realize efficient semantic feature extraction for semantic communications.

In recent years, semantic communications have attracted widespread interest and yielded promising results, among which joint source-channel coding (JSCC) is considered as a key technology\cite{2023magazine_zym,daijinchengjsac}. Typically, a pair of DL-enabled codecs are deployed at the transceivers to generate continuously-valued symbols for analog transmission. In this end-to-end training, source coding and channel coding are no longer separated
, and channel characteristics are introduced into the training process\cite{2019tccn_deepjscc}. 
However, in existing works including audio transmission\cite{wei2022semaudio}, image transmission\cite{2019tccn_deepjscc}, multi-task semantic communications\cite{hu2023scalable}, etc., 
{\color{black} the designs of codec neural network (NN) architecture are focused on. However, the fading channels, which are randomly generated based on the set statistical characteristics, are implicit in the training process.}
These works exploit channel fading by implicitly training, which we call CF-train. The CF-train scheme presents obvious disadvantages when applied to 6G applications, where different sets of parameters need to be trained and stored in smart devices for different fading channels. Nevertheless, channel fading is diverse, and excessive training is not advisable. Therefore, it is important to solve the transfer problem between different channels for semantic communications.

Meanwhile, the orthogonal frequency division multiplexing systems integrating non-orthogonal multiple access technology, i.e., the OFDM-NOMA systems, are recognized as the outstanding multiple access candidates for 6G networks\cite{liu2022NOMA6G}. 
Hence the combination of NOMA and semantic communications attracts attention. \cite{xuxiaodong2023} proposes an asymmetric quantizer to discretize the continuous-valued semantic features for intelligent multi-user detection. In \cite{multimedia2023}, a semi-NOMA method is presented for semantic information multiple access transmission. However, these works ignore the problem of codecs failure caused by channel type changes. In this letter, we propose a channel-transferable semantic communications (CT-SemCom) framework for multi-user OFDM-NOMA systems. To the best of our knowledge, this is the first work to address channel transfer problem in semantic communications.
The main contributions are summarized as follows:
\begin{itemize}
\item {\color{black}A novel CT-SemCom framework is proposed to eliminate the excessive training on various fading channels, which is ignored in existing semantic communications.}
The proposed CT-SemCom framework is first integrated with the multi-user OFDM-NOMA systems.
\item A semantic transmission-oriented power allocation problem is formulated for the OFDM-NOMA systems to realize the transfer from additive white Gaussian noise (AWGN) channels to multi-subcarrier fading channels. To reduce the complexity, a semantics-similar dual transformation (SSDT) algorithm is proposed to obtain analytical solutions of power allocation.
\item Simulation results show the excellent performance of the proposed framework and the SSDT algorithm in channel transferability. Compared with the existing framework, an excess of $4.2-7.3$~dB with respect to peak signal-to-noise ratio (PSNR) is achieved under different variances of Rayleigh fading channels.
\end{itemize}

The rest of this paper is organized as follows. Section~\uppercase\expandafter{\romannumeral2} presents the CT-SemCom framework. Section~\uppercase\expandafter{\romannumeral3} formulates the power allocation problem for the multi-user OFDM-NOMA systems. Section~\uppercase\expandafter{\romannumeral4} proposes the SSDT algorithm with complexity analysis. Section~\uppercase\expandafter{\romannumeral5} evaluates the performance of the proposed framework and the SSDT algorithm. Finally, Section~\uppercase\expandafter{\romannumeral6} draws the conclusion.
\section{Channel-Transferable Semantic Communications Framework}

\begin{figure}
  \centering
  \includegraphics[width=0.47\textwidth]{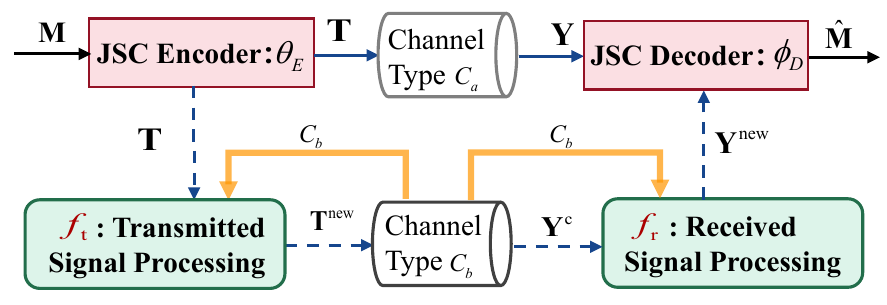}
  \caption{Channel-transferable semantic communications framework.}
  \label{Framework}
\end{figure}

In our proposed CT-SemCom framework, the parameters of codecs learned on one type of channel are expected to apply on other channels. As shown in Fig.~\ref{Framework}, the transfer of different channels can be realized by processing the transmitted signals and received signals with new channel information integrated, where the details are described as follows.

{\color{black}The proposed framework consists of two procedures, i.e., the training process and the testing process. Firstly, JSCC technology is utilized in the codecs training process.}
Semantic features are extracted from the input messages $\mathbf{M}$ by joint source-channel (JSC) encoder $\theta_E$, and the transmitted signals are generated, which are expressed as $\mathbf{T}=\theta_{E}(\mathbf{M})$.
Then via type ${C}_a$ channel, the received signals $\mathbf{Y}$ are input to the JSC decoder $\phi_D$ for message reconstructions, which are denoted by $\hat{ \mathbf{M}}=\phi_{D}(\mathbf{Y})$.
The encoder $\theta_E$ and decoder $\phi_D$ are trained to minimize the loss function
\begin{align}
    \mathcal{L}(\theta_{E},\phi_D) = \Vert \hat{\mathbf{M}}-\mathbf{M} \Vert_2^2,
\end{align}
and then the optimized parameters for type $C_a$ channel will be obtained, which are denoted by  $\Gamma^*=[\theta_{E}^*,\phi_D^*]$.

{\color{black}Furthermore, for the testing process of the proposed CT-SemCom framework, the set of optimized parameters $\Gamma^*$ is deployed to assist signal processing as a fixed semantic encoding and decoding method.}
When the type of channel transfers from $C_a$ to $C_b$, the transmitted signals and received signals are processed as $\mathbf{T}^\text{new}=f_\text{t}(\mathbf{T},C_b)$ and $\mathbf{Y}^\text{new}=f_\text{r}(\mathbf{Y}^\text{c},C_b)$,
respectively, to ensure the quality of semantic communications.
To adapt the received signals $\mathbf{Y^\text{new}}$ to the trained parameters $\Gamma^*$, $\mathbf{Y^\text{new}}$ are expected to be as equal to $\mathbf{Y}$ as possible. Ultimately, the objective of the CT-SemCom framework is to minimize the error between the received signals $\mathbf{Y^\text{new}}$ and $\mathbf{Y}$, which can be formulated as
\begin{equation}\label{idea}
\begin{array}{ll}
\min \quad \!\!\!\!\Vert \mathbf{Y^\text{new}}-\mathbf{Y} \Vert _2^2.
\end{array}
\end{equation}

{\color{black}The proposed CT-SemCom framework can integrate other existing communication systems with different channels to demonstrate its advantages. Next, the integration with OFDM-NOMA systems and the corresponding implementation of channel transfer are detailed.}

\section{System Model and Problem Formulation}

\subsection{OFDM-NOMA System Model Integrated with CT-SemCom}
\begin{figure*}
  \centering
  \includegraphics[width=0.94\textwidth]{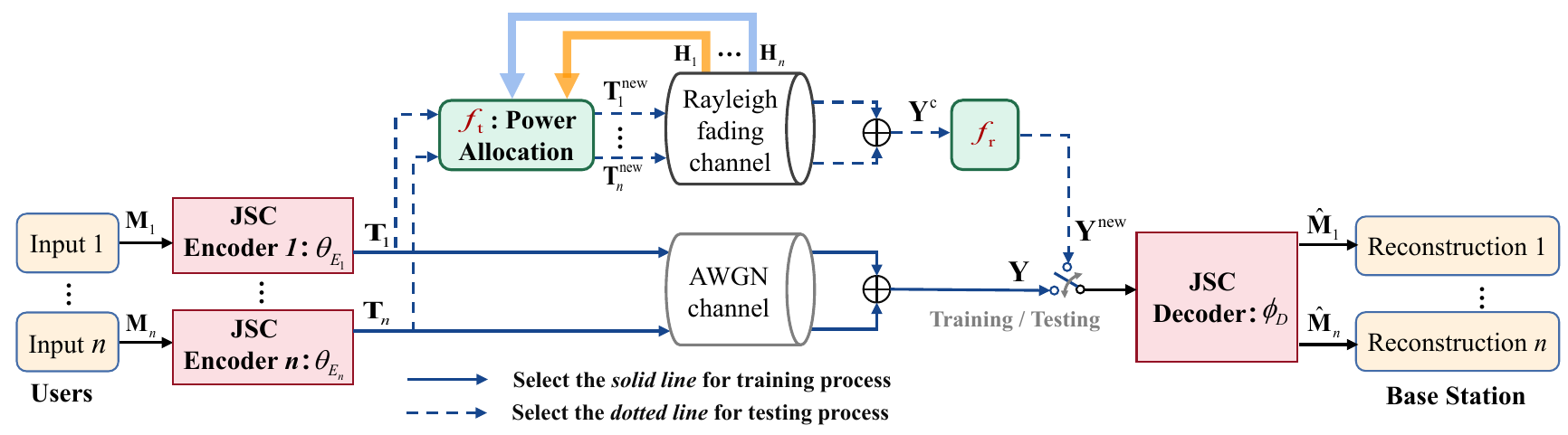}
  \caption{Multi-user OFDM-NOMA systems integrated with CT-SemCom framework.}
  \label{frameNOMA}
\end{figure*}
This letter considers the uplink OFDM-NOMA systems integrated with the proposed CT-SemCom framework, in which the signal processing $f_\text{t}$ and $f_\text{r}$ is specifically designed to realize the transfer from the AWGN channel to multi-subcarrier fading channels. That is, the optimized parameters are obtained by training on the AWGN channel first, and then utilized on fading channels.
As shown in Fig.~\ref{frameNOMA}, multiple {\color{black}non-orthogonal} users denoted by $\mathcal{N}=\{1,2,\cdot\cdot\cdot,N\}$ transmit images to the base station (BS). 
Each user is deployed with a JSC encoder, while the BS is deployed with a JSC decoder.
Thus, $N$ encoders and a single decoder are comprised in the OFDM-NOMA semantic communications systems. {\color{black}While providing services for $U$ users, this system will allocate them into $U/N$ groups, and the signals between each group remain non-interfering.}
{\color{black} Specifically, the NN architecture\footnote[1]{{\color{black}Note that the same NN architecture is employed on both the proposed CT-SemCom framework and the CF-train scheme.}} of the JSC codecs refers to an attention mechanism-based JSCC NN architecture in \cite{snradaptive2021jscc} to adapt to different noise levels.}

The semantic features extracted from the transmitted image $\mathbf{M}_n$ of user~$n$ by JSC encoder $\theta_{E_n}$ are denoted by $\mathbf{X}_n=\theta_{E_n}(\mathbf{M}_n)\in \mathbb{C}^{L\times K}$, where $K$ is the number of subcarriers, $L$ is the number of OFDM-NOMA symbols, and a single symbol carries $K$ features for transmission. $x_{n,l,k}\in \mathbf{X}_n$ denotes the feature of user~$n$ at the $k$-th subcarrier of the $l$-th OFDM-NOMA symbol. 
Then the semantic features $\mathbf{X}_{n,l}$ of each user's OFDM-NOMA symbol at all subcarriers are normalized to transmitted signals $\mathbf{T}_{n,l}$, which are subject to each user's own average power constraint $P_n$, and represented as 
\begin{align}
    \mathbf{T}_{n,l}=\frac{\sqrt{P_n}}{\Vert \mathbf{X}_{n,l} \Vert_{2}}\mathbf{X}_{n,l}\in \mathbb{C}^{K}.
\end{align}

Via the AWGN channel, the received multi-user superimposed signals at BS can be expressed as
\begin{equation}\label{super_signals}
    \mathbf{Y}=\sum\nolimits^N_{n=1} \!\!\mathbf{T}_n + \mathbf{W}\in \mathbb{C}^{L\times K},
\end{equation}
where $\mathbf{T}_n=\{\mathbf{T}_{n,1},\cdot\cdot\cdot, \mathbf{T}_{n,L}\}$, $\mathbf{W}\in \mathbb{C}^{L\times K}$ denotes noise matrix, the element of which is independent and identically distributed complex Gaussian noise term with variance $\sigma_\text{e}^2$, and is denoted by $w_{l,k}\sim \mathcal{CN}(0,\sigma_\text{e}^2)$.
Then the JSC decoder $\phi_D$ deployed at receiver BS reconstructs the received superimposed signals $\mathbf{Y}$ into the transmitted images, which are indicated as $\mathbf{\hat{M}}=\phi_D(\mathbf{Y}) \in \mathbb{C}^{L\times K \times N}$,
where $\mathbf{\hat{M}}=[\mathbf{\hat{M}}_1,\mathbf{\hat{M}}_2,\cdot\cdot\cdot, \mathbf{\hat{M}}_N]$ is the concatenation of all users' reconstructed images.
The loss function is defined as 
\begin{align}
    \mathcal{L}(\theta_{E_1},\cdot\cdot\cdot,\theta_{E_N},\phi_D) = \sum\nolimits_{n=1}^N  \!\!\left(\Vert \mathbf{\hat{M}}_n-\mathbf{M}_n \Vert_2^2\right),
\end{align}
and the optimized parameters obtained through training are denoted by $\Gamma^*=[\theta_{E_1}^*,\cdot\cdot\cdot,\theta_{E_N}^*,\phi_D^*]$.

While applying (or testing) the optimized parameters $\Gamma^*$ on the multi-subcarrier fading channels, the signal processing $f_\text{t}(\cdot)$ specifically refers to power allocation in this OFDM-NOMA system. The power allocation function $f_\text{t}(\cdot)$ is designed to adapt $\mathbf{T}_n$ to the new channel, hence the newly transmitted signals are denoted by 
 \begin{align}\label{f_t}
     \mathbf{T}_n^\text{new}=f_\text{t}(\mathbf{T}_n,\mathbf{H}),
 \end{align}
where $\mathbf{T}_n^\text{new}=\{\mathbf{T}_{n,1}^\text{new},\cdot\cdot\cdot, \mathbf{T}_{n,L}^\text{new}\}$, $\mathbf{H}=[\mathbf{{H}}_1,\mathbf{{H}}_2, \cdot\cdot\cdot, \mathbf{{H}}_N]\in \mathbb{C}^{L\times K\times N}$ is channel fading coefficients matrix, each element of which obeys Rayleigh distribution with variance $\sigma_\text{f}^2$. Next, the received superimposed signals $\mathbf{Y^\text{c}}$ are processed to get the new input of decoder $\phi_D$, which are expressed as
\begin{align}\label{f_r}
    \mathbf{Y^\text{new}}=f_\text{r}(\mathbf{Y^\text{c}}) =f_\text{r}(\sum\nolimits_{n=1}^N \mathbf{H}_n \mathbf{T}^\text{new}_n + \mathbf{W}^\text{new}),
\end{align}
where the element $w_{l,k}^\text{new}$ of $\mathbf{W}^\text{new}$ has the same distribution as $w_{l,k}\in \mathbf{W}$, and is denoted by $w_{l,k}^\text{new}\sim \mathcal{CN}(0,\sigma_\text{e}^2)$.
The average signal-to-noise ratio (SNR) of user~$n$ is defined as $SNR_n=10\text{log}_{10} \frac{P_n}{\sigma_\text{e}^2}$~dB to reflect the noise level at BS.

\subsection{Semantic Transmission-Oriented Power Allocation Problem Formulation}
For the OFDM-NOMA system integrated with the proposed CT-SemCom framework, the objective is to minimize $d_0=\Vert \mathbf{Y^\text{new}}-\mathbf{Y} \Vert _2^2$. 
To further clarify the power allocation problem, we expand the real and imaginary parts of complex signals and channels, which are summarized in Table~\ref{notations}. Therefore, function $f_\text{t}(\cdot)$ in \eqref{f_t} is re-expressed as
 \begin{equation}
\begin{array}{ll}
t^\text{new}_{n,l,k}=t^\text{r}_{n,l,k}p^\text{r}_{n,l,k}+i t^\text{i}_{n,l,k}p^\text{i}_{n,l,k}  \in \mathbf{T}^\text{new}_n,  
\end{array}    
\end{equation}
where $p_{n,l,k}^{\mathrm{r}}$ and $ p_{n,l,k}^{\mathrm{i}}$ are the results of power allocation towards $t_{n,l,k}^{\mathrm{r}}$ and $ t_{n,l,k}^{\mathrm{i}}$, respectively.
\begin{table}
\centering
{
\caption{NOTATIONS}\label{notations}
\renewcommand\arraystretch{1.1}
\scalebox{0.95}{
\begin{tabular}{|c|c|c|c|}
\hline \!\!Matrix\!\! & Element  & \!\!Matrix\!\! & Element \\
\hline $\mathbf{T}_n$ & \!$t_{n,l,k}\!=\!t^\text{r}_{n,l,k}\!+\!i t^\text{i}_{n,l,k}$ &  $\mathbf{W}$ & $w_{l,k}\!=\!w^\text{r}_{l,k}\!+\!i w^\text{i}_{l,k}$ \\
\hline $\mathbf{T}^\text{new}_n$ &\!\! $t^\text{new}_{n,l,k}\!\!=\!t^\text{newr}_{n,l,k}\!+\!i t^\text{newi}_{n,l,k}$ &\!$\mathbf{W}^\text{new}$ & $w^\text{new}_{l,k}\!=\!w^\text{newr}_{l,k}\!+\!i w^\text{newi}_{l,k}$\\
\hline $\mathbf{Y}$ & $y_{l,k}\!=\!y^\text{r}_{l,k}\!+\!i y^\text{i}_{l,k}$ & $\mathbf{Y}^\text{c}$ & $y^\text{c}_{l,k}\!=\!y^\text{cr}_{l,k}\!+\!i y^\text{ci}_{l,k}$\\
\hline $\mathbf{Y}^\text{new}$ & $y^\text{new}_{l,k}\!=\!y^\text{newr}_{l,k}\!+\!i y^\text{newi}_{l,k}$&$\mathbf{H}_n$ & \!\!$h_{n,l,k}\!\!=\!h^\text{r}_{n,l,k}\!\!+\!i h^\text{i}_{n,l,k}$\\
\hline
\end{tabular}}}
\end{table}
A power factor $\alpha$ is defined for the function $f_\text{r}(\cdot)$ in \eqref{f_r}, which is used to adjust the signal amplitude at the receiver BS and re-expressed as ${y^\text{new}_{l,k}}=\alpha {y^\text{c}_{l,k}}$.
The objective function $d_0$ is re-expressed as 
\begin{align}\label{d1}
\begin{array}{ll}
\!\!\!\!d_1 &\!\!\!\!= \sum\nolimits_{l=1}^L\!\sum\nolimits_{k=1}^K\left( \vert y^{\mathrm{new}}_{l,k}\!-\! y_{l,k}\vert ^2\right)\\
&\!\!\!\!=\sum\nolimits_{l=1}^L\!\sum\nolimits_{k=1}^K\!\!\left(\!\left(\alpha y^{\mathrm{cr}}_{l,k}\!-\! y_{l,k}^\text{r}\right)^2\!
+\!\left(\alpha y^{\mathrm{ci}}_{l,k}\!-\! y_{l,k}^\text{i}\right)^2\right),
\end{array}
\end{align}
where $y_{l,k}^\text{r}=\sum_{n=1}^N t_{n, l,k}^{\mathrm{r}}+w_{l,k}^{\mathrm{r}}$, $y_{l,k}^\text{i}=\sum_{n=1}^N t_{n, l,k}^{\mathrm{i}}+w_{l,k}^{\mathrm{i}}$, $y^{\mathrm{cr}}_{l,k}=\sum\nolimits_{n=1}^N\!\left(t_{n, l,k}^{\mathrm{r}} p_{n, l,k}^{\mathrm{r}} h_{n, l,k}^{\mathrm{r}}\!-t_{n, l,k}^{\mathrm{i}} p_{n, l,k}^{\mathrm{i}} h_{n, l,k}^{\mathrm{i}}\right)\!+w_{l,k}^{\mathrm{newr}}$, and 
$y^{\mathrm{ci}}_{l,k}\!=\!\sum\nolimits_{n=1}^N\!\left(\!t_{n, l,k}^{\mathrm{r}} p_{n, l,k}^{\mathrm{r}} h_{n, l,k}^{\mathrm{i}}\!\!+\!t_{n, l,k}^{\mathrm{i}} p_{n, l,k}^{\mathrm{i}} h_{n, l,k}^{\mathrm{r}}\!\right)\!+\!w_{l,k}^{\mathrm{newi}}$.

To sum up, the channel-transferable power allocation problem for the OFDM-NOMA systems is formulated as
\begin{equation}
\begin{array}{ccl}
(\textbf{P1}) &\!\!\!\!\!\! \min\limits_{\begin{subarray}{c}
  p_{n,l,k}^{\mathrm{r}}, p_{n,l,k}^{\mathrm{i}},\alpha   \end{subarray}} & d_1\\
&\!\!\!\!\!\!\text { s.t. } &\!\!\!\!\!\!\!\!\!\!\!\mathcal{C}_1\!: \!\!\frac{1}{K} \!\sum\limits_{k=1}^K \!\!\left(\!\!\left(\!t_{n, l,k}^{\mathrm{r}} p_{n, l,k}^{\mathrm{r}}\!\right)^2\!\!\!+\!\!\left(\!t_{n, l,k}^{\mathrm{i}} p_{n, l,k}^{\mathrm{i}}\!\right)^2\!\right)\!\! \leq \!\!P_n,\\
& &\quad\quad\quad\quad\quad\quad\quad \forall{n\in\mathcal{N}},\forall{l\in\mathcal{L}},\\
\end{array}
\end{equation}
where $P_n$\footnote[1]{\label{PnAWGN}Note that the transmission power is normalized to ${P_n}$ during the training process, i.e., $\Vert \mathbf{T}_{n,l} \Vert_{2}^2=P_n$. Besides, the transmission power does not exceed ${P_n}$ after process $f_\text{t}(\cdot)$ when testing on fading channel, i.e., $\Vert \mathbf{T}^\text{new}_{n,l} \Vert_{2}^2\leq P_n$.}~denotes the maximum average transmission power of user~$n$ at all subcarriers of each OFDM-NOMA symbol. Constraint $\mathcal{C}_1$ ensures that the transmission power is limited.
\section{SSDT Algorithm Design with Low Complexity for Analytical Solutions to Power Allocation} \label{III_details}

\subsection{Analysis of Problem $(\mathbf{P1})$ and Iterative Algorithm}
The noise term $w_{l,k}$ and $w^\text{new}_{l,k}$ of $d_1$ in \eqref{d1} need to be preprocessed according to their distribution.
Then the new objective $d_2$ is derived as
\begin{equation}
\begin{array}{ll}\label{objP1_varience}
d_2 
&\!\!\!\!\!\!=\!\sum\limits_{l=1}^L \!\sum\limits_{k=1}^K\!\left(\!\!E\!\!\left(\begin{array}{l}
\!\!\!\!\left(\alpha\!\left(\dot{y}_{l, k}^{\mathrm{cr}}\!+\!w_{l, k}^{\mathrm{newr}}\!\right)\!\!-\!\!\left(\dot{y}_{l, k}^{\mathrm{r}}\!+\!w_{l, k}^{\mathrm{r}}\!\right)\!\right)^2 \\
\!\!\!\!+\!\!\left(\!\alpha\!\!\left(\!\dot{y}_{l, k}^{\mathrm{ci}}\!+\!w_{l, k}^{\mathrm{newi}}\!\right)\!\!-\!\!\left(\!\dot{y}_{l, k}^{\mathrm{i}}\!+\!w_{l, k}^{\mathrm{i}}\!\right)\!\right)^2
\end{array}\!\!\!\right)\!\!\right) \\
 &\!\!\!\!\!\!\!\!\!\!\!=\! \sum\limits_{l=1}^L \!\sum\limits_{k=1}^K\!\!\left(\!\!\left(\!\!\alpha \dot{y}_{l, k}^{\mathrm{cr}}\!-\!\dot{y}_{l, k}^{\mathrm{r}}\!\right)^2\!\!\!+\!\!\left(\!\alpha \dot{y}_{l, k}^{\mathrm{ci}}\!-\!\dot{y}_{l, k}^{\mathrm{i}}\!\right)^2\!\!\!+\!\!2\!\left(\!\alpha^2\!+\!1\!\right) \!\sigma_\text{e}^2\!\!\right)\!,
\end{array}
\end{equation}
where $\dot{y}$ are the non-noise terms of corresponding ${y}$, for example, ${y}_{l, k}^{\mathrm{cr}}=\dot{y}_{l, k}^{\mathrm{cr}}+w_{l, k}^{\mathrm{newr}}$, {\color{black}$E(\cdot)$ denotes mathematical expectation.} 
Then problem ($\mathbf{P1}$) is equivalently translated to
\begin{equation}
\begin{array}{lcl}
\!\!\!\!\!\!\!\!\!\!\!\!\!\!\!(\textbf{P2}) & \min\limits_{p_{n,l,k}^{\mathrm{r}}, p_{n,l,k}^{\mathrm{i}},\alpha} & d_2\\
&\text { s.t. }& \mathcal{C}_1.\quad
\quad\quad\end{array}
\end{equation}
It can be seen that problem ($\mathbf{P2}$) is a quadratically constrained quadratic programming (QCQP) problem when $\alpha$ is fixed, which can be solved by the interior point method \cite{boyd2004convex}. When $p_{n,l,k}^{\mathrm{r}}$ and $p_{n,l,k}^{\mathrm{i}}$ are fixed, the optimal $\alpha$ is obtained by $\frac{\partial d_2}{\partial \alpha}=0$. Thus, the solutions to problem ($\mathbf{P2}$) are obtained through finite iterations between the power allocation and the power factor. The complexity of the interior point method is approximately $\mathcal{O}\left(L(2NK)^{3}\right)$. Thus the complexity of this iterative algorithm for problem ($\mathbf{P2}$) is $\mathcal{O}\left(I_\text{max}\left(L(2NK)^{3}+1\right)\right)$, where $I_\text{max}$ is the maximum iteration number. 
\subsection{SSDT Algorithm}
Problem ($\mathbf{P2}$) is not conducive to the communications of massive devices since the complexity of the interior point method grows cubically with the dimension of variables. 
With this fact in mind, we perform semantics-similar dual transformation (SSDT) on problem ($\mathbf{P2}$), i.e., exchange the objective and constraints while ensuring semantic transmission performance. 
The SSDT algorithm reduces the complexity by deriving analytical solutions to the transformed problem, which is detailed as follows:
It can be seen from \eqref{objP1_varience} that the power factor $\alpha$ affects both PSNR and transmission power. When the power constraint $\mathcal{C}_1$ is not considered, the first two square terms in \eqref{objP1_varience} are optimized to zero. The smaller $\alpha$ is, the smaller the objective function $d_2$ is, and the better PSNR performance can be achieved. As $\alpha$ becomes smaller, the transmission power must become larger to minimize the first two square terms. To not violate the power constraint $\mathcal{C}_1$, $\alpha$ can not be infinitesimal. 
Since the square terms are important parts affecting PSNR, $\alpha$ is expected to be the smallest under the premise that the power does not exceed the constraint.

 Therefore, the first step of the SSDT algorithm is to quickly get an appropriate $\alpha$ as detailed below, instead of multiple iterations. Next, with the fixed $\alpha$,
problem $(\mathbf{P2})$ is approximately transformed into the following form
\begin{equation}
\begin{array}{cl}
\!\!(\mathbf{P3} )  \!\!\min\limits_{\begin{subarray}{c}
  p_{\!n\!,l\!,k}^{\mathrm{r}}, p_{\!n\!,l\!,k}^{\mathrm{i}}   \end{subarray}} &\!\!\!\!\!\!\!\!p_\text{sum}\!=\!\!\!\!\sum\limits_{n=\!1}^N\!\! \beta\!_n \!\!\sum\limits_{l=\!1}^L \!\sum\limits_{k=\!1}^K\!\!\left(\!\!\left(t_{n, l, k}^{\mathrm{r}} p_{n, l, k}^{\mathrm{r}}\right)^2\!\!\!+\!\!\left(\!t_{n, l, k}^{\mathrm{i}} p_{n, l, k}^{\mathrm{i}}\!\right)^2\!\right)\quad \quad\\
\quad \quad\text { s.t. }
&  \!\!\mathcal{C}_2:(\alpha \dot{y}_{l, k}^{\mathrm{cr}}-\dot{y}_{l, k}^{\mathrm{r}})^2=0, \forall k \in \mathcal{K}, \forall l \in \mathcal{L}, \\
&  \!\!\mathcal{C}_3:(\alpha \dot{y}_{l, k}^{\mathrm{ci}}-\dot{y}_{l, k}^{\mathrm{i}})^2=0, \forall k \in \mathcal{K}, \forall l \in \mathcal{L}.
\end{array}
\end{equation}
Problem $(\mathbf{P3})$ is a weighted sum-power minimization problem. $\beta_n$ is a ratio factor to balance transmission power of each user. Constrains $\mathcal{C}_2$ and $\mathcal{C}_3$ aim to guarantee that the objective function $d_2$ of problem $(\mathbf{P2})$ is minimized. Although the transformed problem $(\mathbf{P3})$ is not exactly equivalent to problem ($\mathbf{P2}$), it can approach the performance of problem ($\mathbf{P2}$), which will be exhibited in Sec.~V.


Problem  $(\mathbf{P3})$ can be dealt with by the Lagrangian dual method. The Lagrangian function is given by
\begin{equation}
\begin{array}{ll}
&L(p_{n, l, k}^{\mathrm{r}}, p_{n, l, k}^{\mathrm{i}}, \lambda_{l, k}, \mu_{l, k})\\
&=p_\text{sum}+\lambda_{l, k}(\alpha \dot{y}_{l, k}^{\mathrm{cr}}-\dot{y}_{l, k}^{\mathrm{r}})+\mu_{l, k}(\alpha \dot{y}_{l, k}^{\mathrm{ci}}-\dot{y}_{l, k}^{\mathrm{i}})
\end{array}
\end{equation}
where $\lambda_{l, k}$ and $\mu_{l, k}$ are the Lagrangian multipliers associated with the  constrains $\mathcal{C}_2$ and $\mathcal{C}_3$, respectively. We derive partial derivatives of the Lagrangian function on variables. By solving simultaneous equations $\frac{\partial L}{\partial p_{n, l, k}^{\mathrm{r}}}=0$, $\frac{\partial L}{\partial p_{n, l, k}^{\mathrm{i}}}=0$, $\frac{\partial L}{\partial \lambda_{l, k}}=0$, and $\frac{\partial L}{\partial \mu_{l, k}}=0$, the power allocation results are expressed as
\begin{equation}\label{pnr}
 p_{n, l, k}^{\mathrm{r}}=-\frac{\alpha (\lambda_{l, k} h_{n, l, k}^{\mathrm{r}}+ \mu_{l, k} h_{n, l, k}^{\mathrm{i}})}{2 \beta_n t_{n, l, k}^{\mathrm{r}}},
\end{equation}
\begin{equation}\label{pni}
 p_{n, l, k}^{\mathrm{i}}=\frac{\alpha (\lambda_{l, k} h_{n, l, k}^{\mathrm{i}}- \mu_{l, k} h_{n, l, k}^{\mathrm{r}})}{2 \beta_n t_{n, l, k}^{\mathrm{i}}},
\end{equation}
where 

$\lambda_{l, k}\!=\!\frac{-2 \sum_{n=1}^N t_{n, l, k}^{\mathrm{r}}}{ \alpha^2\!\! \sum\limits_{n=1}^N \!\!\! \frac{\left(h_{n, l, k}^{\mathrm{r}}\right)^2\!+\!\left(h_{n, l, k}^{\mathrm{i}}\right)^2}{\beta_n}},$
$\mu_{l, k}\!=\!\frac{-2 \sum_{n=1}^N t_{n, l, k}^{\mathrm{i}}}{\alpha^2 \!\!\sum\limits_{n=1}^N \!\!\! \frac{\left(h_{n, l, k}^{\mathrm{r}}\right)^2\!+\!\left(h_{n, l, k}^{\mathrm{i}}\right)^2}{\beta_n}}.$

According to \eqref{pnr} and \eqref{pni}, $\beta_n=1/\sqrt{P_n}$ is defined to approximately meet the proportion of each user's power limit. We calculate the transmission power of each user's each symbol $p_{n,l} \!\!=\!\! \sum\nolimits_{k=1}^K\!\!\left(\!(t_{n, l, k}^{\mathrm{r}} p_{n, l, k}^{\mathrm{r}})^2\!\!+\!\!(t_{n, l, k}^{\mathrm{i}} p_{n, l, k}^{\mathrm{i}})^2\!\right)$ based on \eqref{pnr}-\eqref{pni}, and then make each $p_{n,l}$ reach the maximum power $P_n$ to obtain the corresponding $\alpha_{n,l}$ derived as
\begin{equation}\label{alpha_suboptimal}
\alpha_{n,l}\!\!=\!\!\!\!\!\sqrt{\!\sum_{k=1}^K\! \!\frac{\!\!\left(\!\!\left(\!h_{n, l, k}^{\mathrm{r}}\!\right)^2\!\!\!+\!\!\left(\!h_{n, l, k}^{\mathrm{i}}\!\right)^2\!\right)\!\!\left(\!\!\left(\!\sum\limits_{n=\!1}^N \!t_{n, l, k}^{\mathrm{r}}\!\!\right)^2\!\!\!+\!\!\left(\!\sum\limits_{n=\!1}^N t_{n, l, k}^{\mathrm{i}}\!\!\right)^2\!\right)}{K P_n\left(\beta_n\right)^2\left(\sum\limits_{n=1}^N \frac{\left(h_{n, l, k}^{\mathrm{r}}\right)^2+\left(h_{n, l, k}^{\mathrm{i}}\right)^2}{\beta_n}\right)^2}}.
\end{equation}
$\alpha\!=\!\max\{\alpha_{1,1}, \cdot\!\cdot\!\cdot, \alpha_{n,l},\cdot\!\cdot\!\cdot,\alpha_{N,L}\}$ is determined to make all $p_{n,l}$ satisfy their own power constraints. {\color{black} The complexity of calculating $\alpha_{n,l}$ and finding the maximum $\alpha$ of the unordered array $\alpha_{n,l}$ is $\mathcal{O}\left(2NL\right)$. The complexity of the derived analytical solutions is $\mathcal{O}\left(2NLK\!+\!2LK\right)$. Thus the complexity of SSDT algorithm is analyzed as $\mathcal{O}\left(2L(NK\!+\!K\!+\!N)\right)$, which is significantly lower than the complexity of problem $(\mathbf{P2})$.}


\section{Simulation Results}
In this section, simulation results are exhibited to demonstrate the performance of the proposed framework. The simulation parameters and experimental settings are as follows: {\color{black}$N$ non-orthogonal users} perform image transmission based on $K$ = 64 subcarriers. Cifar-10 dataset is employed for demonstration. {\color{black} The Adam is employed to perform backpropagation with $\beta_1$ = 0.9, $\beta_2$ = 0.99 during the training process, which takes 450 epochs, including 200 epochs with a learning rate of 5e-4 and 250 epochs with a learning rate of 5e-6.}
{\color{black} The above training parameters are employed on both the proposed framework and the CF-train scheme. The performance of CF-train shown below is trained under variance of Rayleigh fading $\sigma_\text{f}^2=3$~dB.}
The transmission power is limited to $P_1=0.8$~w, $P_2=0.2$~w, and $P_3=0.2$~w. The noise level at the receiver is reflected by $SNR_1=10\text{log}_{10} \frac{P_1}{\sigma_\text{e}^2}$~dB.

\begin{figure}
  \centering
  \includegraphics[width=0.45\textwidth]{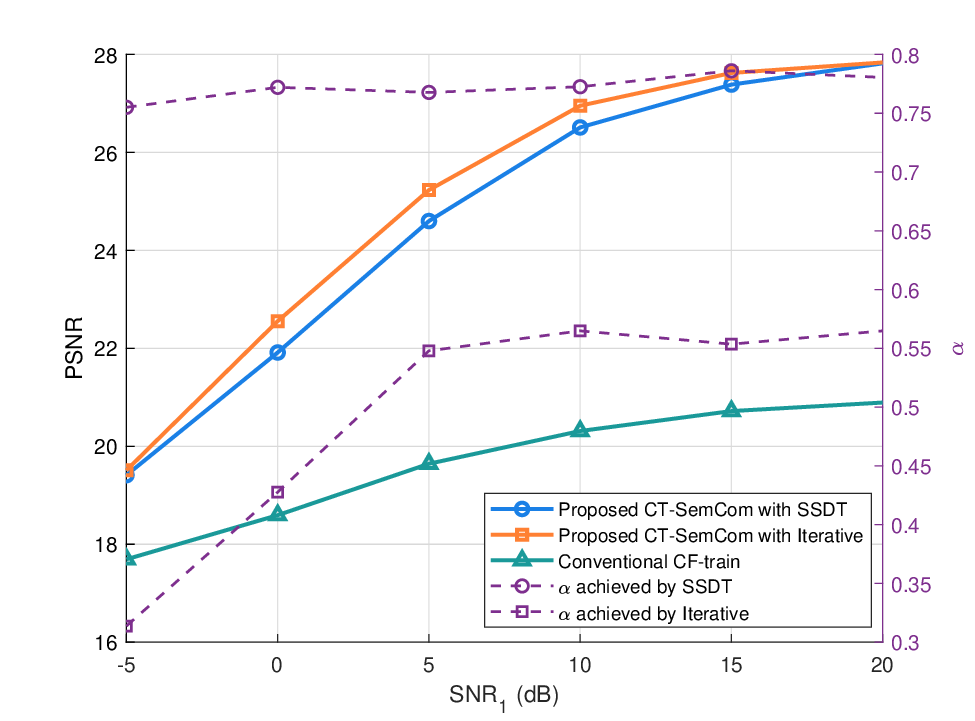}
  \caption{Average PSNR and optimized power factor $\alpha$ among different schemes as $SNR_1$ increases.}
  \label{SNR_PSNR}
 \end{figure}
 
Fig.~\ref{SNR_PSNR} shows the mean PSNR performance and optimized power factor $\alpha$ among different schemes as $SNR_1$ increases under variance of Rayleigh fading $\sigma_\text{f}^2=3$~dB and  {\color{black}$N=2$}.
Both the proposed SSDT algorithm and iterative algorithm based on the proposed CT-SemCom framework outperform the conventional CF-train scheme w.r.t. PSNR, where the SSDT algorithm exceeds the CF-train scheme 1.7 -- 6.9~dB for different $SNR_1$. It illustrates that the proposed CT-SemCom framework achieves good transfer from AWGN channels to Rayleigh fading channels. Besides, the factor $\alpha$ obtained by the SSDT algorithm is larger than the iterative algorithm, which echoes that the smaller $\alpha$ is, the better PSNR is achieved. Moreover, {\color{black}the average runtime of SSDT algorithm and the iteraive algorithm is $0.0021$~s and $11.7776$~s, respectively.} The PSNR of the proposed SSDT algorithm only degrades $1.40\%$ with a large degree of complexity reduction.
 \begin{figure}
  \centering
  \includegraphics[width=0.45\textwidth]{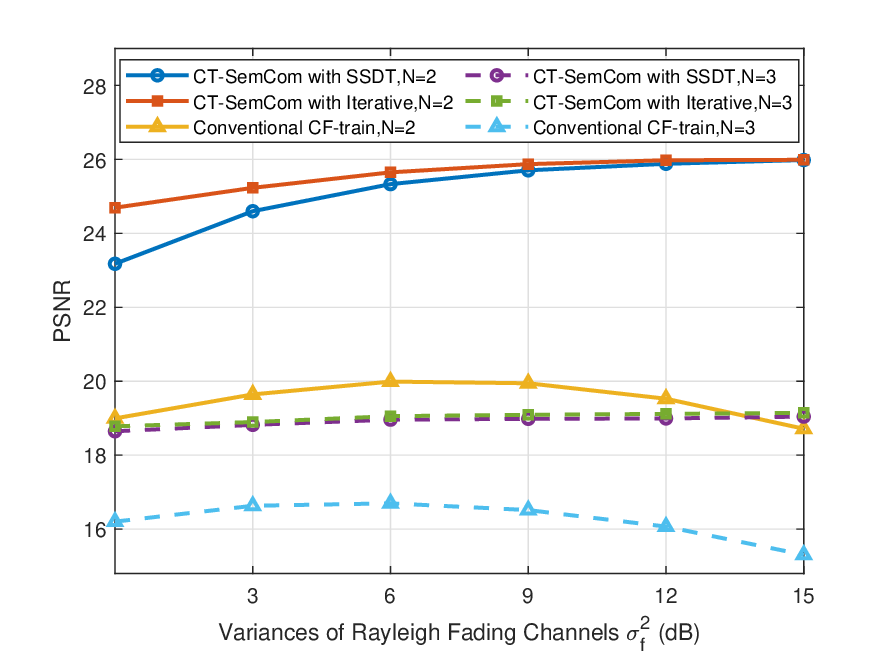}
  \caption{Average PSNR under different $N$ changes with variances of Rayleigh fading channels $\sigma_\text{f}^2$.}
  \label{fading_sigma_psnr}
 \end{figure}

{\color{black}Fig.~\ref{fading_sigma_psnr} illustrates PSNR performance of different number of non-orthogonal users as variances of Rayleigh fading channels $\sigma_\text{f}^2$ changes under $SNR_1=5$~dB.}
The SSDT algorithm based on the proposed CT-SemCom framework outperforms the CF-train scheme by 4.2~\textendash~7.3~dB in PSNR among different variances when $N=2$. {\color{black}The PSNR decreases as the number of non-orthogonal users increases, due to the increased difficulty in decoding superposition signals.} Moreover, the PSNR of the CF-train scheme becomes worse as the variances $\sigma_\text{f}^2$ changes. However, the PSNR of the proposed SSDT algorithm and iterative algorithm maintain stable, which illustrates that the transfer from AWGN channels to Rayleigh fading channels with different statistical characteristics is realized and the proposed framework achieves excellent channel transferability.

\section{Conclusion}
In this letter, a novel CT-SemCom framework is proposed to eliminate the excessive training on different channels. Integrated with the multi-user OFDM-NOMA systems, a semantic transmission-oriented power allocation problem is solved to realize the transfer from AWGN channels to fading channels.
The SSDT algorithm is proposed, which significantly reduces the computational complexity with only slight performance drops.
Simulation results verify the superiority of the proposed framework, which not only achieves good channel transferability performance from AWGN channels to different Rayleigh fading channels, \!but also significantly outperforms the conventional training scheme.

\ifCLASSOPTIONcaptionsoff
  \newpage
\fi

\bibliographystyle{IEEEtran}
\bibliography{refersSecNOMA}

\begin{thebibliography}{10}
\providecommand{\url}[1]{#1}
\csname url@samestyle\endcsname
\providecommand{\newblock}{\relax}
\providecommand{\bibinfo}[2]{#2}
\providecommand{\BIBentrySTDinterwordspacing}{\spaceskip=0pt\relax}
\providecommand{\BIBentryALTinterwordstretchfactor}{4}
\providecommand{\BIBentryALTinterwordspacing}{\spaceskip=\fontdimen2\font plus
\BIBentryALTinterwordstretchfactor\fontdimen3\font minus
  \fontdimen4\font\relax}
\providecommand{\BIBforeignlanguage}[2]{{%
\expandafter\ifx\csname l@#1\endcsname\relax
\typeout{** WARNING: IEEEtran.bst: No hyphenation pattern has been}%
\typeout{** loaded for the language `#1'. Using the pattern for}%
\typeout{** the default language instead.}%
\else
\language=\csname l@#1\endcsname
\fi
#2}}
\providecommand{\BIBdecl}{\relax}
\BIBdecl

\bibitem{intelligent_apps}
K.~B. Letaief, W.~Chen, Y.~Shi, J.~Zhang, and Y.-J.~A. Zhang, ``The roadmap to
  {6G}: {AI} empowered wireless networks,'' \emph{{IEEE} Commun. Mag.},
  vol.~57, no.~8, pp. 84--90, 2019.

\bibitem{engineering_ping}
P.~Zhang, W.~Xu, H.~Gao, K.~Niu, X.~Xu, X.~Qin, C.~Yuan, Z.~Qin, H.~Zhao,
  J.~Wei \emph{et~al.}, ``Toward wisdom-evolutionary and primitive-concise
  {6G}: A new paradigm of semantic communication networks,''
  \emph{Engineering}, vol.~8, pp. 60--73, 2022.

\bibitem{2023magazine_zym}
W.~Xu, Y.~Zhang, F.~Wang, Z.~Qin, C.~Liu, and P.~Zhang, ``Semantic
  communication for the {I}nternet of vehicles: A multiuser cooperative
  approach,'' \emph{{IEEE} Veh. Technol. Mag.}, 2023.

\bibitem{daijinchengjsac}
J.~Dai, S.~Wang, K.~Tan, Z.~Si, X.~Qin, K.~Niu, and P.~Zhang, ``Nonlinear
  transform source-channel coding for semantic communications,'' \emph{{IEEE}
  J. Sel. Areas Commun.}, vol.~40, no.~8, pp. 2300--2316, 2022.

\bibitem{2019tccn_deepjscc}
E.~Bourtsoulatze, D.~B. Kurka, and D.~G{\"u}nd{\"u}z, ``Deep joint
  source-channel coding for wireless image transmission,'' \emph{{IEEE} Trans.
  on Cogn. Commun. Netw.}, vol.~5, no.~3, pp. 567--579, 2019.

\bibitem{wei2022semaudio}
H.~Wei, W.~Xu, F.~Wang, X.~Du, T.~Zhang, and P.~Zhang, ``Sem{A}udio:
  Semantic-aware streaming communications for real-time audio transmission,''
  in \emph{Proc. IEEE Global Communications Conference (GLOBECOM)}.\hskip 1em
  plus 0.5em minus 0.4em\relax IEEE, 2022, pp. 3965--3970.

\bibitem{hu2023scalable}
J.~Hu, F.~Wang, W.~Xu, H.~Gao, and P.~Zhang, ``Scalable multi-task semantic
  communication system with feature importance ranking,'' in \emph{Proc. IEEE
  International Conference on Acoustics, Speech and Signal Processing
  (ICASSP)}.\hskip 1em plus 0.5em minus 0.4em\relax IEEE, 2023, pp. 1--5.

\bibitem{liu2022NOMA6G}
Y.~Liu, S.~Zhang, X.~Mu, Z.~Ding, R.~Schober, N.~Al-Dhahir, E.~Hossain, and
  X.~Shen, ``Evolution of {NOMA} toward next generation multiple access
  ({NGMA}) for {6G},'' \emph{{IEEE} J. Sel. Areas Commun.}, vol.~40, no.~4, pp.
  1037--1071, 2022.

\bibitem{xuxiaodong2023}
W.~Li, H.~Liang, C.~Dong, X.~Xu, P.~Zhang, and K.~Liu, ``Non-orthogonal
  multiple access enhanced multi-user semantic communication,'' \emph{arXiv
  preprint arXiv:2303.06597}, 2023.

\bibitem{multimedia2023}
Y.~Duan, Q.~Du, X.~Fang, Z.~Xie, Z.~Qin, X.~Tao, C.~Pan, and G.~Liu,
  ``Multimedia semantic communications: Representation, encoding and
  transmission,'' \emph{{IEEE} Netw.}, vol.~37, no.~1, pp. 44--50, 2023.

\bibitem{snradaptive2021jscc}
X.~Bao, M.~Jiang, and H.~Zhang, ``{ADJSCC}-l: {SNR}-adaptive {JSCC} networks
  for multi-layer wireless image transmission,'' in \emph{Prco. International
  Conference on Computer and Communications (ICCC)}.\hskip 1em plus 0.5em minus
  0.4em\relax IEEE, 2021, pp. 1812--1816.

\bibitem{boyd2004convex}
S.~Boyd, S.~P. Boyd, and L.~Vandenberghe, \emph{Convex optimization}.\hskip 1em
  plus 0.5em minus 0.4em\relax Cambridge university press, 2004.

\end{thebibliography}



\end{document}